\documentclass[ApJL,twocolumn]{aastex62}

\usepackage{amssymb}
\usepackage{amsmath}
\usepackage{graphicx}
\usepackage{dcolumn}
\usepackage{hyperref}
\usepackage{color,units}
\usepackage{aas_macros}
\usepackage{lineno}
\usepackage{xspace}
\usepackage{subfigure}
\usepackage{comment}
\usepackage{newtxtext,newtxmath}

\usepackage{xcolor}

\newcommand{\SPA}{School of Physics and Astronomy, Monash University, Clayton VIC 3800, Australia}
\newcommand{\OzGravMonash}{OzGrav: The ARC Centre of Excellence for Gravitational Wave Discovery, Clayton VIC 3800, Australia}

\shorttitle{Heavy double neutron stars}
\shortauthors{Galaudage et al.}

\begin{document}

\title{Heavy Double Neutron Stars: Birth, Midlife, and Death}

\author{Shanika Galaudage}
\email{shanika.galaudage@monash.edu}
\affiliation{\SPA}
\affiliation{\OzGravMonash}

\author{Christian Adamcewicz}
\affiliation{\SPA}
\affiliation{\OzGravMonash}

\author{Xing-Jiang Zhu}
\affiliation{\SPA}
\affiliation{\OzGravMonash}

\author{Simon Stevenson}
\affiliation{Centre for Astrophysics and Supercomputing, Swinburne University of Technology, Hawthorn, VIC 3122, Australia}
\affiliation{OzGrav: The ARC Centre of Excellence for Gravitational Wave Discovery, Hawthorn VIC 3122, Australia}

\author{Eric Thrane}
\affiliation{\SPA}
\affiliation{\OzGravMonash}

\begin{abstract}
Radio pulsar observations probe the lives of Galactic double neutron star (DNS) systems while gravitational waves enable us to study extragalactic DNS in their final moments.
By combining measurements from radio and gravitational-wave astronomy, we seek to gain a more complete understanding of DNS from formation to merger.
We analyse the recent gravitational-wave binary neutron star mergers GW170817 and GW190425 in the context of other DNS known from radio astronomy.
By employing a model for the birth and evolution of DNS, we measure the mass distribution of DNS at birth, at midlife (in the radio), and at death (in gravitational waves).
We consider the hypothesis that the high-mass gravitational-wave event GW190425 is part of a subpopulation formed through unstable case BB mass transfer, which quickly merge in $\sim~10-100~\mathrm{Myr}$.
We find mild evidence to support this hypothesis and that GW190425 is not a clear outlier from the radio population as previously claimed.
If there are fast-merging binaries, we estimate that they constitute $8-79\%$ of DNS at birth (90\% credibility).
We estimate the typical delay time between the birth and death of fast-merging binaries to be $\approx 5-401~\mathrm{Myr}$ (90\% credibility).
We discuss the implications for radio and gravitational-wave astronomy.
\end{abstract}

\section{Introduction}
Radio and gravitational-wave astronomy enable us to study double neutron stars (DNS) at different stages of binary evolution.
Radio observations, which probe DNS at midlife, have so far provided precise mass measurements for 12 Galactic DNS \citep[see, e.g., Table 1 in][and references therein]{Zhu2020}.
In the standard formation scenario, a DNS consists of two neutron stars: a first-born ``recycled'' neutron star sped up from accretion and a second-born ``slow'' neutron star.
With relatively few measurements, it is not yet possible to precisely measure the distribution of DNS component masses, but current observations of Galactic DNS are consistent with the following hypotheses: the recycled neutron star mass is distributed according to a double-Gaussian distribution with peaks at $\mu_1\approx 1.34~{M_\odot}, \mu_2\approx 1.47~{M_\odot}$ while the slow neutron star mass is distributed uniformly on the interval $1.16-1.42~{M_\odot}$~\citep{Farrow2019}.

Based on these fits, one may conclude that most double neutron stars are relatively low-mass: 95\% of recycled neutron stars have masses less than $1.64~{M_\odot}$ while 95\% of slow neutron stars have masses less than $1.43~{M_\odot}$.
Curiously, pulsars in binaries with white dwarfs are regularly found with much larger masses; for example, the mass of PSR J0740+6620 is measured to be $\sim 2.14~{M_\odot}$~{\citep{Cromartie2020}}.
Population synthesis studies have explored the DNS populations in the Milky Way, modelling the Galactic population of double neutron stars and understanding the difference between the radio and gravitational-wave populations \citep{Vigna-Gomez2018, Chattopadhyay2020, Kruckow2020, Mandel2021}.  

Gravitational-wave astronomy probes only the final moments of DNS.
To date, LIGO--Virgo \citep{AdLIGO,Virgo} have reported the discovery of two DNS mergers: GW170817~\citep{GW170817} and GW190425~\citep{GW190425}.
While the component masses of GW170817 are consistent with the masses of Galactic DNS, \citet{GW190425} argue it is difficult to explain GW190425, with a primary mass of $1.60{M_\odot} - 1.87{M_\odot}$ (90\% credibility, low-spin prior), as a member of the Galactic radio population.
This difficulty has led to speculation about the origin of GW190425.

\cite{RomeroShaw2020} argue that GW190425 is best understood as a fast-merging binary, which underwent unstable case BB mass transfer.
They speculate that GW190425 could have formed from a progenitor including a $4-5 M_\odot$ He star with a $\sim 3 M_\odot$ CO core capable of producing a $\sim 2 M_\odot$ neutron star~\citep{Mueller16,ErtlWoosley20}.
According to this hypothesis, massive DNS like GW190425 are not observed in the radio because they merge before they are likely to be seen.
This fast-merging hypothesis seems to rely on the premise that the most massive slow-merging DNS are disrupted by large supernova kicks.
In the appendix, we present the results of a population synthesis study to show that large kicks $\gtrsim\unit[2000]{km\, s^{-1}}$ are required for this theory to have explanatory power.

Meanwhile, \cite{Safarzadeh2020} argue that binaries like GW190425 are common in the Milky Way, but we do not see them because the magnetic field of such massive binaries is buried through accretion.
\cite{GW190425} also floats the hypothesis that GW190425 could have been dynamically assembled, but this seems unlikely; numerical simulations suggest NS tend to be expelled from the cores of dense stellar environment leading to low rates of dynamical mergers~\citep{Zevin2019,Grindlay2016,Bae2014,Belczynski2018,Ye2020,Papenfort2018}.\footnote{\cite{Andrews2019} posit that there are observational hints of dynamically assembled DNS, despite pessimistic predictions from numerical studies.}
A range of arguably more exotic hypotheses have been put forward as well, for example, that the primary object in GW190425 could be a black hole \citep{Kyutoku2020,Clesse2020,Gupta2020,Foley2020}.

In order to understand how GW190425 formed, it is necessary to introduce a unified framework, which simultaneously accounts for radio and gravitational-wave observations.
Selection effects are likely to make the radio-visible and gravitational-wave populations disparate.
However, the two populations are linked by the population of DNS \textit{at birth}.
By modeling the population properties of DNS at birth, and then evolving the population through midlife to death, we simultaneously fit data from radio and gravitational-wave astronomy.
The remainder of this paper is organised as follows.
In Section~\ref{sec:method}, we introduce a formalism for estimating the birth properties of DNS using measurements from both gravitational waves and radio.
In Section~\ref{sec:results}, we apply our analysis to the two LIGO--Virgo DNS mergers and the 12 DNS in order to estimate the population properties of DNS at birth, midlife, and death.
Finally, in Section~\ref{sec:discussion}, we discuss the implications of our results and future work.

\section{Method}\label{sec:method}
We outline a method of inferring the birth mass distributions from observations of radio-visible DNS and binary neutron star mergers.
In section \ref{subsec:birth_model}, we introduce an observationally motivated model \citep[based on][]{Farrow2019} for the mass distribution of the DNS population at birth.
We describe models for the evolution of the DNS formation rate density within the Milky Way (Galactic) and in the greater universe (extragalactic).
In section \ref{subsec:evolve}, we describe how to evolve the birth mass distribution in time to obtain the radio-visible and gravitational-wave populations.
We describe our model for the distribution of delay time between binary formation and merger.
In section \ref{subsec:inferparams}, we describe a formalism to infer the population properties of DNS from data. 

\subsection{Modeling the DNS population at birth}\label{subsec:birth_model}
We parameterise the mass distribution of DNS at birth, building on mass models from \cite{Farrow2019}.
These models explored the radio-visible mass distribution for slow and recycled neutron star.
For simplicity, we assume the recycled and slow neutron star mass distributions are independent.
We take the model used for radio-visible DNS as a starting point to describe the DNS mass distribution at birth.
\cite{Farrow2019} provides parameterised fits for the distributions of slow and recycled neutron stars in DNS.
Their best-fit model for slow neutron stars was a uniform distribution, but we opt to use a Gaussian, which is also consistent with radio data.

Our model for the mass distribution of slow neutron stars $\pi(m_s)$ is a double Gaussian,
\begin{align}\label{eq:slow_two_gaussian}
     \pi(m_{s} |\xi_s, \mu_{si}, &\sigma_{si})= &  \xi_s \mathcal{N}(\mu_{s1},\sigma_{s1}) + (1 - \xi_s)\mathcal{N}(\mu_{s2},\sigma_{s2}),
\end{align}
where the distribution $\mathcal{N}(\mu_{s1},\sigma_{s1})$ is a normalized Gaussian distribution for the low-mass peak with mean $\mu_{s1}$ and width $\sigma_{s1}$ and $\mathcal{N}(\mu_{s2},\sigma_{s2})$ is a normalized Gaussian distribution for the high-mass peak with mean $\mu_{s2}$ and width $\sigma_{s2}$.
The subscripts 1 and 2 refer to the low and high mass peaks respectively.
We argue below that that the radio population can be understood as coming from peak $1$ while the gravitational-wave population includes both peak $1$ and $2$.

The high-mass peak is an addition to the model put forward in~\cite{Farrow2019}.
This peak is motivated by GW190425 and the existence of a high-mass peak at $\sim 1.8~M_{\odot}$ in the mass distribution of neutron stars in binaries with white dwarfs \citep{Kiziltan2013, Alsing2018}.
Motivated by the arguments put forth in \citep{RomeroShaw2020}, we posit that this peak is associated with a fast-merging DNS population formed through a stage of unstable case BB mass transfer\footnote{While the stable case BB mass transfer process was capable of explaining observations prior to GW190425, the unstable mass transfer hypothesis was invoked to reconcile GW190425 and radio DNS observations.}.
In this scenario, the DNS progenitor undergoes a second phase of common envelope evolution which involves a neutron star and a helium star with a carbon-oxygen core. At the end of this evolution, the helium envelope gets ejected, resulting in a tight binary \citep{Ivanova2003,Dewi2003}.
It is hypothesized that high-mass (second-born) neutron stars are preferentially associated with this scenario because the binary is likely to be disrupted (during the second supernova) without being in a sufficiently tight orbit (assuming large supernova kicks for high-mass neutron stars).
We note that the recycling process usually refers to  stable case BB mass transfer. For the unstable mass transfer that leads to the second common envelope described above, the exact amount of accretion is not well understood; \citet{MacLeod2014} found an estimate of $0.05-0.1~M_{\odot}$ but \citet{Tauris17dns} adopted an upper limit of $0.01~M_{\odot}$.

\begin{figure*}
    \centering
    \includegraphics[width = 0.48\textwidth]{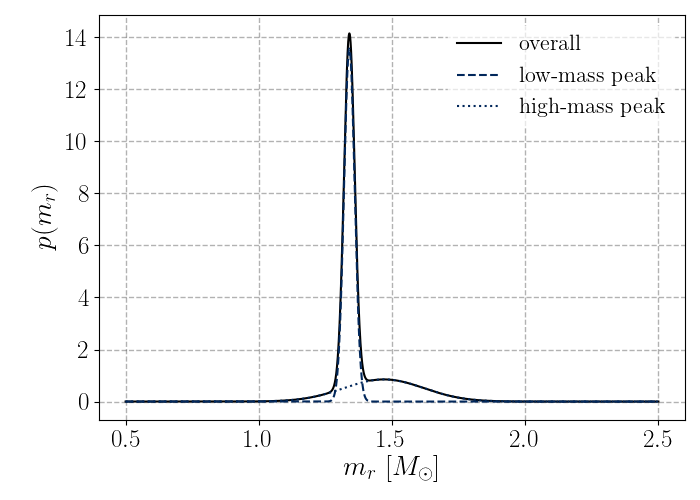}
    \hspace{5mm}
    \includegraphics[width = 0.48\textwidth]{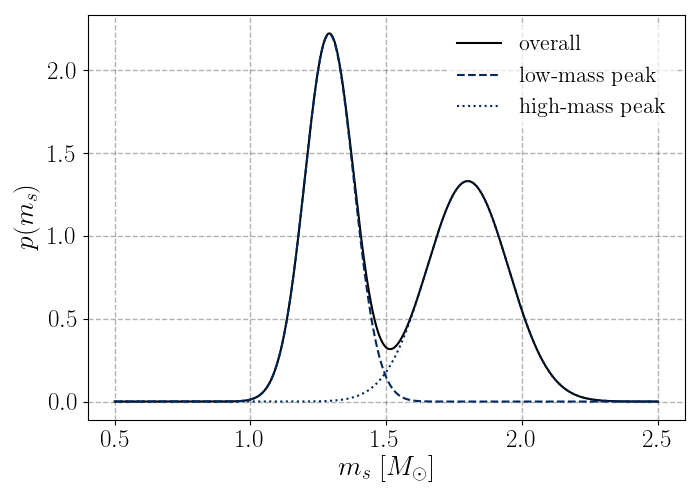}
    \caption{Example birth mass distributions for recycled (left) and slow (right) neutron stars assuming the fiducial population parameters given in Table~\ref{tab:fiducial}.
    The dashed line traces the low-mass peak, the dotted line traces the high-mass peak, and the solid line is the overall distribution.
    The high-mass peak in the distribution of slow neutron stars corresponds to fast-merging binaries, which we posit are not frequently seen in radio due to selection effects.
    }
    \label{fig:mrms}
\end{figure*}

For the recycled distribution we use the preferred model from \cite{Farrow2019}:
\begin{align}\label{eq:recycled_two_gaussian}
     \pi(m_{r} |\xi_r, \mu_{ri}, & \sigma_{ri})= 
     &  \xi_r \mathcal{N}(\mu_{r1},\sigma_{r1}) + (1 - \xi_r)\mathcal{N}(\mu_{r2},\sigma_{r2}),
\end{align}
where the distribution $\mathcal{N}(\mu_{r1},\sigma_{r1})$ is a normalized Gaussian distribution for the low-mass peak with mean $\mu_{r1}$ and width $\sigma_{r1}$ and $\mathcal{N}(\mu_{r2},\sigma_{r2})$ is a normalized Gaussian distribution for the high-mass peak with mean $\mu_{r2}$ and width $\sigma_{r2}$.
Note that the high-mass peak in the recycled distribution is not related to the fast-merging channel associated with the high-mass peak in the slow visible DNS population; it is neutron star population feature of the recycled radio-visible distribution.
Example plots of mass distributions of recycled and slow neutron stars are shown in Fig.~\ref{fig:mrms}.
This fiducial population is characterised by parameters provided in Table~\ref{tab:fiducial}.
We return to this fiducial distribution throughout as an illustrative example.

\begin{table}
\begin{tabular}{lp{3.5cm}cc} 
    \hline
    Parameter & Description & Recycled & Slow \\
    \hline\hline
    $\xi$ & Fraction of binaries in the low-mass peak & 0.68 & 0.5 \\
    $\mu_\text{1}$ & Mean of low-mass peak & 1.34 $M_{\odot}$ & 1.29 $M_{\odot}$ \\
    $\sigma_\text{1}$ & Width of low-mass peak & 0.02 $M_{\odot}$ & 0.09 $M_{\odot}$\\
    $\mu_\text{2}$ & Mean of high-mass peak & 1.47 $M_{\odot}$ & 1.8 $M_{\odot}$\\
    $\sigma_\text{2}$ & Width of high-mass peak & 0.15 $M_{\odot}$ & 0.15 $M_{\odot}$\\
    \hline
\end{tabular}
\caption{
    Fiducial population parameters describing an example population of double neutron stars at birth.
    See Eqs.~\ref{eq:slow_two_gaussian}-\ref{eq:recycled_two_gaussian} for parameter definitions.
    The population is plotted in Fig.~\ref{fig:mrms}.
}
\label{tab:fiducial}
\end{table}

In order to understand how the DNS mass distribution evolves in time we need to model the DNS birth rate density $R_b$ over cosmic time.
We use two models for birth rate density: one for the DNS formation rate in the Milky Way, and another for the extragalactic population.
For the DNS formation rate in the Milky Way, we model the rate density as constant over cosmic time; see \cite{Snaith2015} Fig. 7.
We assume the extragalactic DNS formation rate tracks the \cite{Madau&Dickinson} model for comoving volumetric star formation rate density as a function of redshift
\begin{align}\label{eq:madau}
    R_b(t) \propto \psi(z(t)) = 0.015\frac{(1 + z(t))^{2.7}}{1 + [(1 + z(t))/2.9]^{5.6}},
\end{align}
where $\psi(z)$ is in units of ${M_{\odot}\mathrm{yr^{-1}Mpc^{-3}}}$.
Moving forward, we distinguish between the Galactic and extragalactic DNS populations, carrying out parallel calculations (outlined in section~\ref{subsec:evolve}) for each population, but assuming different birth rate densities.

\subsection{Evolving the DNS population: midlife and death}\label{subsec:evolve}
Using the birth mass distribution and the birth rate density we can evolve the population in time.
The DNS merger rate density (which is a snapshot of the gravitational-wave population) is given by the convolution of the birth rate density with the delay time distribution:
\begin{align}\label{eq:Rm}
    R_m(t) = \int_0^t dt_b \, R_b(t_b) \pi(t-t_b) .
\end{align}
Here, $R_m(t)$ is the merger rate density as a function of cosmic time while $R_b(t)$ is the DNS formation rate density.
We measure time $t$ relative to the beginning of the binary formation $\approx 13~\mathrm{Gyr}$ ago, which coincides roughly with the origin of the Galaxy at $t=0$.
The variable $t_b$ is the birth time of the binary and $\pi(t-t_b)$ is the probability density function for the delay time between birth and merger.

\begin{figure*}
    \centering
    \includegraphics[width = 0.49\textwidth]{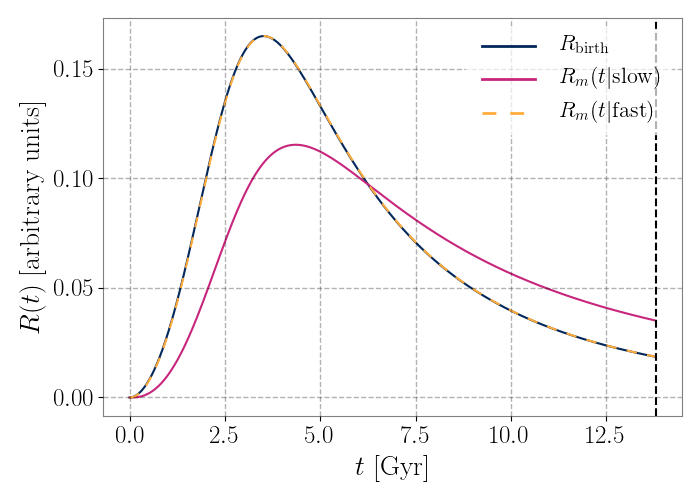}
    \hspace{1mm}
    \includegraphics[width = 0.49\textwidth]{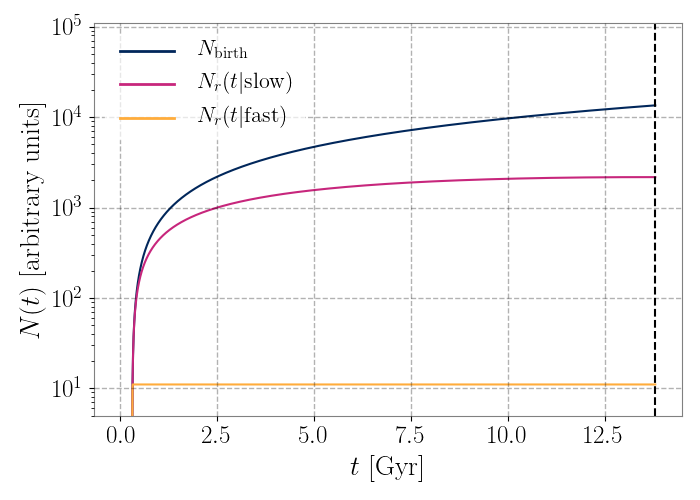}
    \caption{Comparison of the merger rate densities and number of radio-visible binaries in the slow-merging channel and the fast-merging channel. 
    \textit{Left}: Merger rate density (arbitrary units) for different delay-time channels for extragalactic population assuming the Madau-Dickinson model for the DNS formation rate density.
    \textit{Right}: Number of binaries visible in radio (arbitrary units) for different delay-time channels for the Milky Way population assuming a uniform distribution for the DNS formation rate density.
    The black dashed lines represent $t_\text{today}$.
    }
    \label{fig:delay-time-comparison}
\end{figure*}

We employ two models for the time delay distribution: one for conventional (slow) mergers and one for fast mergers.
The conventional model assumes a distribution that is uniform in the logarithm of time delay with a minimum delay time of $\unit[30]{Myr}$:
\begin{align}\label{eq:td_model1}
    \pi(t_d) \propto 
    \begin{cases}
        0 & t_d \leq \unit[30]{Myr} \\
        1/t_d  & t_d > \unit[30]{Myr}
    \end{cases} ;
\end{align}
see, for example, Fig.~3 in~\cite{Neijssel}.
The second model is motivated by our desire to accommodate potentially fast-merging binaries such as GW190425.
We use a delta function distribution for the delay time distribution of the fast-merging channel,  
\begin{align}\label{eq:td_model2}
     \pi(t_{d}) = \delta(t_d - t_*) .
\end{align}
This choice is motivated by the necessity to enforce that this secondary channel is, in fact, fast-merging.
A wide time delay distribution does not place such a constraint on the secondary channel, as DNS under these circumstances are still free to merge on a similar timescale to the standard channel.
Preliminary testing found that a Gaussian distribution of time delay produced results nearly identical to that of a delta function distribution.
As such, we opt for the former to simplify our model.
As in \cite{RomeroShaw2020}, we assume a fiducial fast-merger time of $t_*=10~\mathrm{Myr}$~\citep{GW190425,RomeroShaw2020}.
However, we also present results where $t_*$ is treated as a free parameter, allowed to take on values from $5$ to $500~\mathrm{Myr}$.
Using our prescription for formation rate density and delay-time distribution we can obtain the merger rate density distribution. 
The left panel of Fig.~\ref{fig:delay-time-comparison} shows the DNS formation rate density and the merger rate densities for the slow (Eq. \ref{eq:td_model1}) and fast (Eq. \ref{eq:td_model2}) merging channels; 
in these plots we assume $t_* = 10~\mathrm{Myr}$. These rate densities are in arbitrary units. 

Now let us consider the radio-visible DNS population.
Using the DNS formation and merging rate densities we can calculate the number of radio-bright binaries in the Galaxy given by
\begin{align}\label{eq:Nr}
    N_r(t) = 
    \int_0^t dt' \, 
    \Big(
    R_b(t') - R_m(t') 
    \Big) \epsilon(t') .
\end{align}
Here, $\epsilon$ describes the fraction of midlife DNS in the Galaxy which are radio-visible.
If we temporarily ignore the $\epsilon$ factor for a moment (setting it to one), the number of radio-bright binaries is simply given by the number that have been born over the interval $(0,t)$ minus the number that have merged, and so which are no longer visible.
However, not every yet-to-merge binary is visible in radio.
Something like 10\% of neutron stars are visible because they are not beamed toward the Earth~\citep{richard,lisa_ska}, and so we expect $\epsilon(t)\approx 0.10$.
Moreover, it is possible that $\epsilon$ is time-dependent if we expect pulsars to grow dim as they age \citep{AshleyYuri17,bns_spin}.
For the sake of simplicity, we ignore this possibility for the time being, though, this time-dependence is worth exploring in future work.

The right side panel of Fig.~\ref{fig:delay-time-comparison} shows the number of DNS at birth (navy), the number of radio-visible DNS in the slow channel (magenta), and the number of radio-visible DNS in the fast-merging channel (yellow).
The distribution of fast-merging binaries has a lower number of radio binaries compared to the slow-merging binaries.
In practice, there are significant uncertainties associated with $R_b(t)$ and $\epsilon(t)$. 
However, if we consider two different models, slow-merging and fast-merging for the delay time distribution, and if we know the ratio of the current merger rate densities for these two channels
\begin{align}
    \frac{R_m(t|\text{slow})}{R_m(t|\text{fast})} ,
    \label{eq:merger_rate_ratio}
\end{align}
then we can precisely predict the ratio of currently visible radio binaries for these two models 
\begin{align}
    \frac{N_r(t|\text{slow})}{N_r(t|\text{fast})} .
    \label{eq:radio_number_ratio}
\end{align}
The overall normalisations of $R_b$ and $\epsilon$ drop out of the ratio.

\begin{figure*}
    \centering
    \includegraphics[width = 0.49\textwidth]{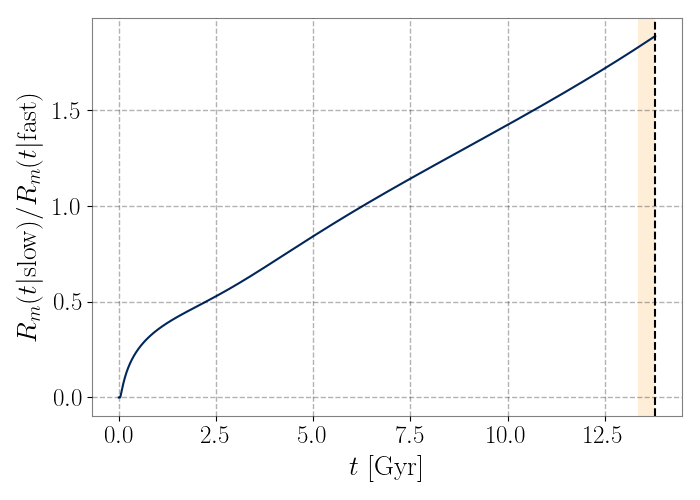}
    \hspace{1mm}
    \includegraphics[width = 0.49\textwidth]{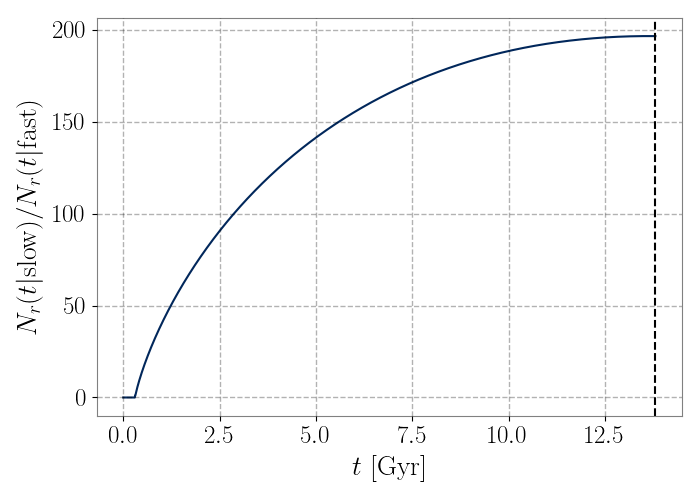}
    \caption{The ratio of the merger rate and number of radio-visible binaries in the slow-merging channel and the fast-merging channel. 
    \textit{Left}: Merger rate ratio of the binaries in different delay-time channels for extragalactic population assuming the Madau-Dickinson model for the DNS formation rate density.
    The shaded region (yellow) represents the detection horizon for gravitational-wave mergers.
    \textit{Right}: Ratio of number of binaries visible in radio for different delay-time channels for the Milky Way population assuming a uniform distribution for the DNS formation rate density.
    The black dashed lines represent $t_\text{today}$.
    }
    \label{fig:ratio_plots}
\end{figure*}

Figure~\ref{fig:ratio_plots} shows the slow-to-fast ratio of merger rate densities (left) and the slow-to-fast ratio of radio-visible binaries (right).
The ratio of fast-merging DNS to slow-merging DNS today is $1.88$.
However, the radio-visible DNS in the fast-merging channel is highly suppressed; the ratio of slow-merging to fast-merging is $196$.
This illustrates how GW190425 could have been part of a fast-merging population (with a merger rate comparable to slow-merging DNS) while accounting for the lack of comparably massive radio-visible DNS.
We calculate these ratios assuming that LIGO--Virgo (and radio observations) are sensitive to DNS in the local Universe: $z\approx 0$.
However, we note that as gravitational-wave detectors become more sensitive, it will be possible to probe DNS mergers over sufficiently large spans of cosmic time.
Thus, it will eventually become necessary to take into account the evolving merger rate.

\begin{figure}
    \includegraphics[width=\columnwidth]{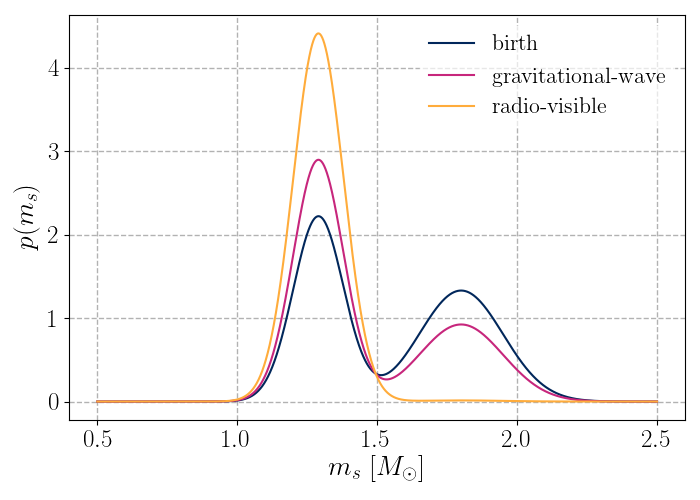}
    \caption{
    The mass distribution of slow neutron stars at birth (blue), merging (magenta), and in radio (yellow).
    The high-mass peak is suppressed in radio by a factor of $\zeta_{\mathrm{radio}}=196$.
    Here, we have assumed that fast-merging binaries merge in $t_* = \unit[10]{Myr}$.
    }
    \label{fig:birth_merged_radio}
\end{figure}

The slow-to-fast merger rate density ratio ($\zeta_{\mathrm{GW}}=1.88$) and the slow-to-fast ratio of radio binaries for ($\zeta_{\mathrm{radio}}=196$) serve as transfer functions that describe how the relative mixture of slow-to-fast evolves from birth to radio binaries to gravitational-wave mergers.
The radio and gravitational-wave distributions of slow neutron stars are given by,
\begin{align}
     \pi_x(m_{s} &|\xi_s, \mu_{si}, \sigma_{si})= \nonumber \\ &  \eta_{x}\zeta_x\xi_s \mathcal{N}(\mu_{s1},\sigma_{s1}) + \eta_{x}(1 - \xi_s)\mathcal{N}(\mu_{s2},\sigma_{s2}),
\label{eq:slow_two_gaussian_today}
\end{align}
where $x=\text{GW}$ employs the gravitational-wave transfer function and $x=\text{radio}$ employs the radio transfer function.
Meanwhile, $\eta$ is a normalisation factor:
\begin{align}
    \eta_x = \frac{1}{(\zeta_x - 1) \xi_s + 1} .
    \label{eq:eta_frac}
\end{align}
where $\zeta_{\mathrm{GW}}$ is the transfer function weight from the merger number ratio and $\zeta_{\mathrm{radio}}$ is the transfer function weight from the radio number ratio; recall that $\xi_s$ is the fraction of slow-merging binaries produced at birth (Eq.~\ref{eq:slow_two_gaussian}).
Figure~\ref{fig:birth_merged_radio} shows the birth DNS distribution for slow mass and the resulting radio and gravitational-wave distributions.

\subsection{Inferring the birth parameters}\label{subsec:inferparams}
Having related the radio and gravitational-wave populations to the birth distribution, we can combine the radio and gravitational-wave data to infer the birth mass distribution.
The likelihood of the data given population parameters $\Lambda$ is the product of the gravitational wave and radio likelihoods:
\begin{align}\label{eq:L_pop}
    {\cal L}(d|\Lambda) = \prod_i^M{\cal L}_\text{GW}(d_{i}|\Lambda)\prod_j^N{\cal L}_\text{radio}(d_{j}|\Lambda) .
\end{align}
The population parameters $\Lambda$ describe the shape of the DNS mass distribution at birth. 
The prior distributions of these parameters are summarised in Table~\ref{tab:parameterpriors}.
We assume a uniform prior $U(0,1)$ for $\xi_s$, the fraction of low neutron stars in the low-mass Gaussian.
Some population studies (e.g., \cite{Safarzadeh2020}) suggest that the fraction of fast-merging binaries is relatively small $\lesssim10\%$, which would correspond to a prior peaked at $\xi_s\approx 90\%$.
However, we take a data-driven approach so that all values of $\xi_s$ are given equal prior weight.

Taking into account selection effects~\citep{Vitale2020,intro} the likelihood is,
\begin{align}\label{eq:L_pop_det}
    {\cal L}(d|\Lambda,\text{det}) = & \left(\frac{1}{p_\text{det,GW}(\Lambda)}\right)^M\prod_i^M{\cal L}_\text{GW}(d_{i}|\Lambda) \nonumber \\ & \left(\frac{1}{p_\text{det,radio}(\Lambda)}\right)^N\prod_j^N{\cal L}_\text{radio}(d_{j}|\Lambda) .
\end{align}
Here, $p_\text{det}$ is the probability of detecting the DNS.
For gravitational events
\begin{align}
p_\text{det, GW} \approx {\cal M}^{5/2} ,
\end{align}
where ${\cal M}$ is the chirp mass,
\begin{align}
    \mathcal{M} = \frac{(m_1 m_2)^{3/5}}{(m_1 + m_2)^{1/5}} ,
\end{align}
and $m_1$ and $m_2$ are the primary and secondary masses of the compact binary respectively.
The recycled and slow masses are related to the primary and secondary masses as follows,
\begin{align}
    m_1 = & \max(m_r, m_s) \\
    m_2 = & \min(m_r, m_s) .
\end{align}

There are two types of selection effects one can imagine at play with the radio population. 
Fast-merging binaries are not visible in radio.
Therefore, the population we see in the radio is not representative of the birth population.
However, we already take this into account with the transfer function $\zeta_\text{radio}$.
Here, we are concerned with how the radio population detectability varies with mass.
We assume that this variation is negligible and so we set $p_\text{det, radio} = 1$.

We re-write the likelihood as a sum that recycles posterior samples; see, e.g., \citep{intro},
\begin{align}\label{eq:L_pop_det_recycled}
    {\cal L}&(d|\Lambda,\text{det}) =  \nonumber \\ & \left(\frac{1}{p_\text{det,GW}(\Lambda)}\right)^M\prod_i^M\frac{
    {\cal L}_\text{GW}(d_i|\Lambda_0)}{n_i}\sum_k^{n_i} \frac{\pi(\theta_{i,k}|\Lambda)}{\pi(\theta_{i,k}|\Lambda_0)} \nonumber \\ & \left(\frac{1}{p_\text{det,radio}(\Lambda)}\right)^N\prod_j^N\frac{{\cal L}_\text{radio}(d_j|\Lambda_0)}{n_j}  \sum_k^{n_j} \frac{\pi(\theta_{j,k}|\Lambda)}{\pi(\theta_{j,k}|\Lambda_0)} .
\end{align}
where $\Lambda_0$ is the fiducial population parameters values used for initial analysis of individual DNS.
We evaluate Eq.~\ref{eq:L_pop_det_recycled} using \textsc{GWPopulation}, a population inference package for gravitational-wave analyses based on \textsc{bilby} \cite{bilby}. 
We use the nested sampler \textsc{dynesty} \citep{dynesty}. 
The recycled and slow mass posteriors for the gravitational-wave events are from analyses by \cite{Zhu2020} and the posteriors for the radio observations are from \cite{Farrow2019}.

\begin{table*}
\centering
\begin{tabular}{lll} 
    \hline
    Parameter & Description & Prior\\
    \hline\hline
    $\xi_{r}$ & Fraction of recycled NS in low-mass Gaussian & $U(0,1)$ \\
    \hline
    $\mu_{r1}$ & Mean of low-mass Gaussian in $m_r$ distribution & $U(1.1M_{\odot},1.5M_{\odot}$) \\
    \hline
    $\sigma_{r1}$ & Standard deviation of low-mass Gaussian in $m_r$ distribution & $U(0.005M_{\odot},0.5M_{\odot}$) \\
    \hline
    $\mu_{r2}$ & Mean of high-mass Gaussian in $m_r$ distribution & $U(\mu_{r1},2.0M_{\odot}$) \\
    \hline
    $\sigma_{r2}$ & Standard deviation of high-mass Gaussian in $m_r$ distribution & $U(0.005M_{\odot},0.5M_{\odot}$) \\
    \hline
    $\xi_{s}$ & Fraction of slow NS in low-mass Gaussian & $U(0,1)$ \\
    \hline
    $\mu_{s1}$ & Mean of low-mass Gaussian in $m_s$ distribution & $U(1.1M_{\odot},1.5M_{\odot}$) \\
    \hline
    $\sigma_{s1}$ & Standard deviation of low-mass Gaussian in $m_s$ distribution & $U(0.005M_{\odot},0.5M_{\odot}$) \\
    \hline
    $\mu_{s2}$ & Mean of high-mass Gaussian in $m_s$ distribution & $U(\mu_{s1},2.0M_{\odot}$) \\
    \hline
    $\sigma_{s2}$ & Standard deviation of high-mass Gaussian in $m_s$ distribution & $U(0.005M_{\odot},0.5M_{\odot}$) \\
    \hline
    $m_\text{min}$ & Minimum possible NS mass & $1M_{\odot}$ \\
    \hline
    $m_\text{max}$ & Maximum possible NS mass & $3M_{\odot}$ \\
    \hline
    $t_{*}$ & characteristic delay time & $U(5~\mathrm{Myr},500~\mathrm{Myr}$)\\
    \hline
\end{tabular}
\caption{
Prior distributions for the hyper-parameters. The notation $U(a,b)$ indicates a uniform distribution on the interval ranging from $a$ to $b$.
}
\label{tab:parameterpriors}
\end{table*}

\section{Results}\label{sec:results}
\begin{figure*}
    \includegraphics[width=\textwidth]{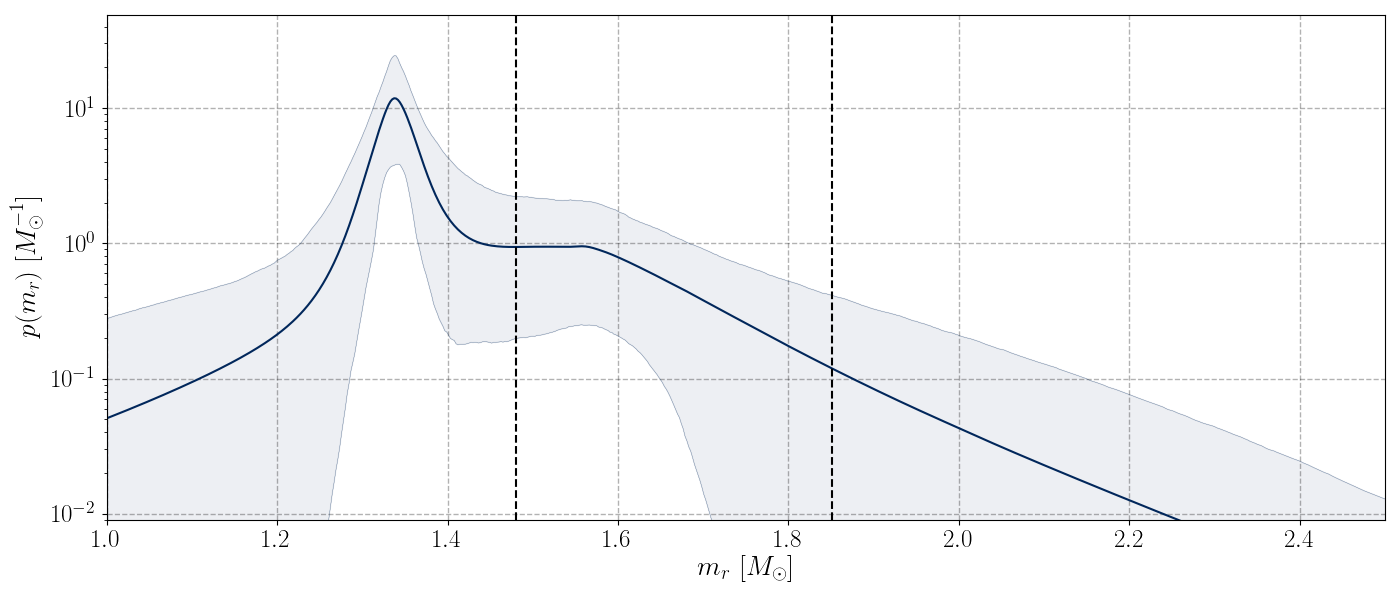}
    \includegraphics[width=\textwidth]{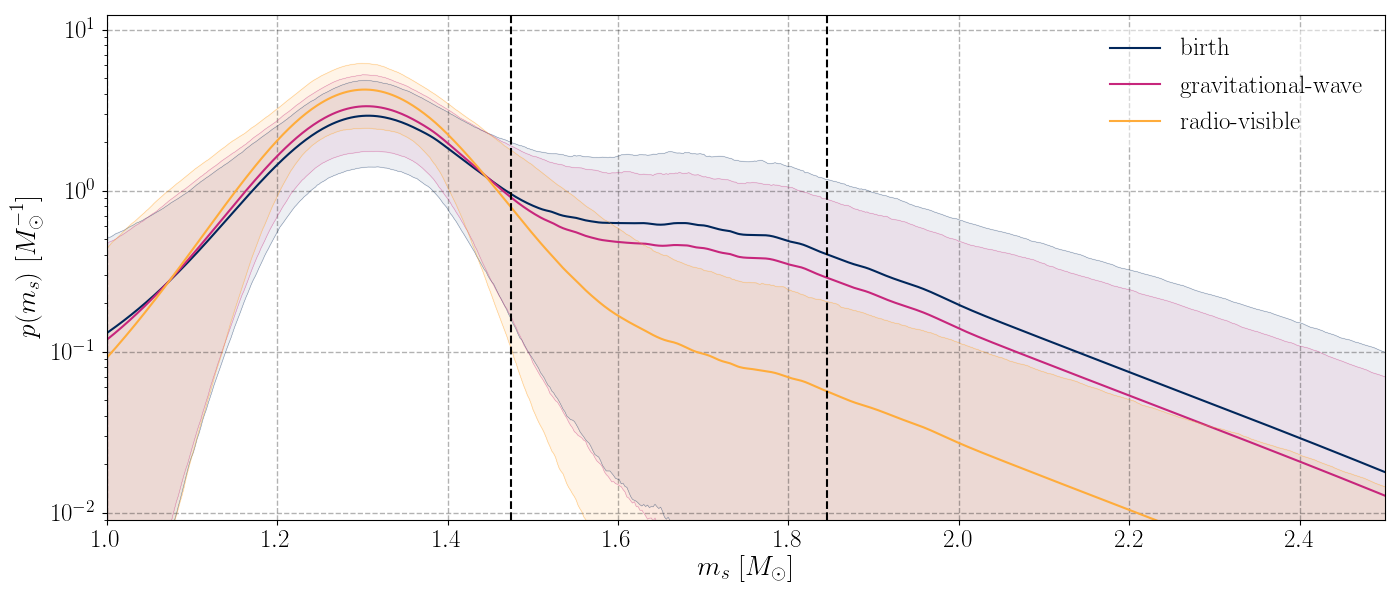}
    \caption{
    Reconstructed distributions for recycled mass ($m_r$; above) and slow mass ($m_s$; below) for the birth (navy), merging (magenta) and radio-visible (yellow) distributions.
    The solid curves indicate the population predictive distributions, and the shaded regions represent the 90\% credible interval.
    The black dashed lines represent the 90\% credible interval of the posterior distribution for GW190425.
    }
    \label{fig:ppd_recycled_slow}
\end{figure*}

We analyse the 12 radio-visible DNS \citep{Ferdman2014,Kramer2006,vanLeeuwen2015,Fonseca2014,Cameron2018,Lynch2018,Weisberg2010,Martinez2015,Ferdman2020,Haniewicz2021,Jacoby2006,Lynch2012} and two binary neutron star mergers from LIGO--Virgo \citep{GW170817, GW190425}.
In Fig.~\ref{fig:ppd_recycled_slow}, we plot the reconstructed recycled and slow neutron-star mass spectra.
The top panel shows the mass distribution of recycled neutron stars and the bottom panel shows the mass distribution of slow neutron stars.
The solid curves are the population predictive distributions (the best-guess reconstruction, averaged over population parameters) and the shaded regions represents the 90\% credible regions.
The birth population is shown in navy.
The merging population observed in gravitational waves is shown in magenta.
The radio-visible population is shown in yellow.
In the mass distribution of recycled neutron stars, the birth, merging and radio distributions are identical. 
In the mass distribution of slow neutron stars, the birth, merging and radio distributions differ substantially.
Fast-merging, high-mass binaries present in the birth distribution are suppressed in the radio distribution (by a factor of {$\zeta_{\mathrm{radio}} \approx 4 - 74$} at 90\% credibility) because they do not live long enough to be observed. 
Fast-merging binaries are suppressed in the merging population by a modest factor of {$\zeta_{\mathrm{GW}}\approx 1.8$} (see Fig.~\ref{fig:transfer_posteriors} for distributions of the transfer function weights).

\begin{figure*}
    \centering
    \includegraphics[width=0.75\textwidth]{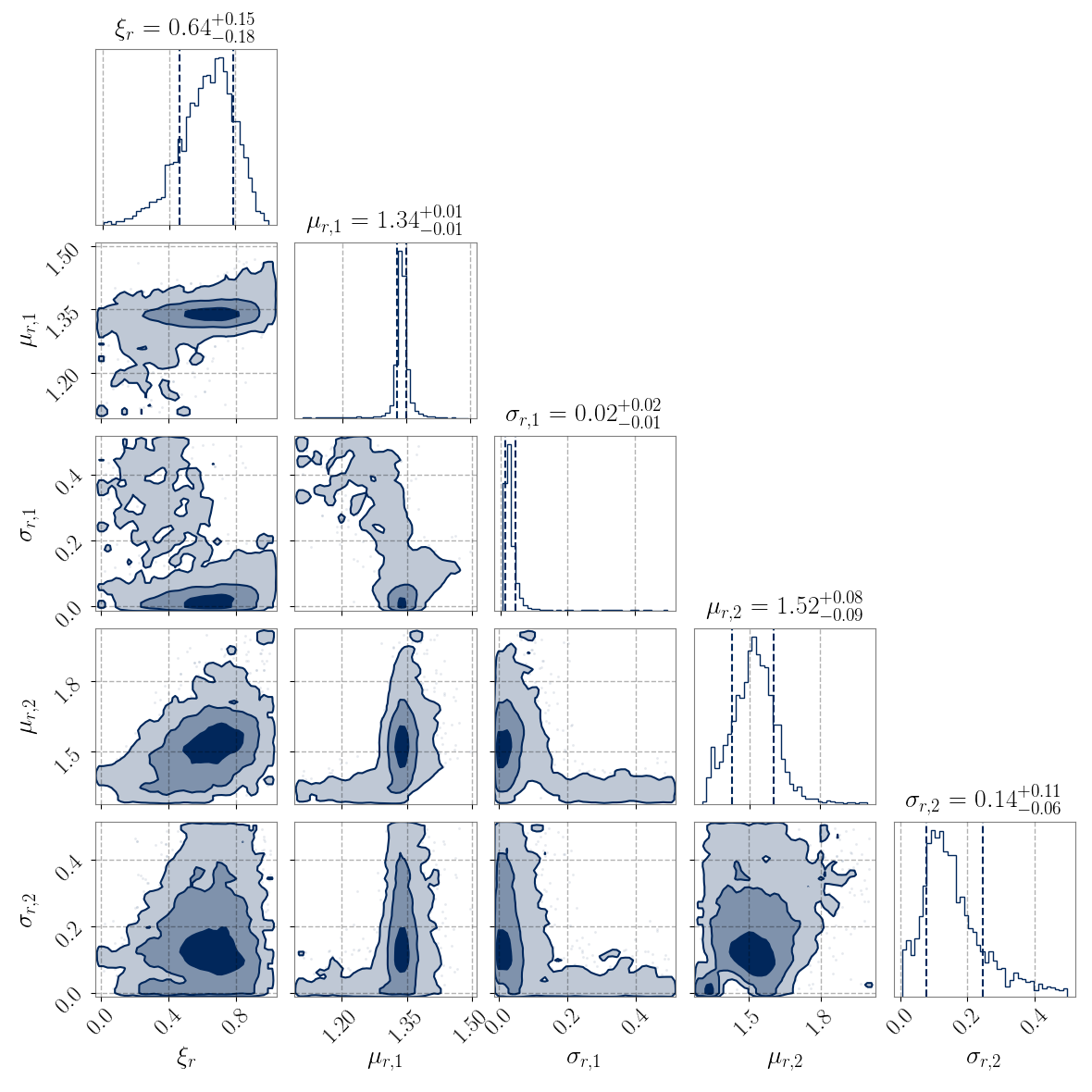}
    \caption{
    Posterior distributions for the recycled mass population parameters.
    The credible intervals on the 2D posterior distributions are at $1\sigma$, $2\sigma$ and $3\sigma$, using increasingly light shading.
    The intervals on the 1D posterior distributions are at $1\sigma$.
    }
    \label{fig:mr}
\end{figure*}

\begin{figure*}
    \centering
    \includegraphics[width=0.75\textwidth]{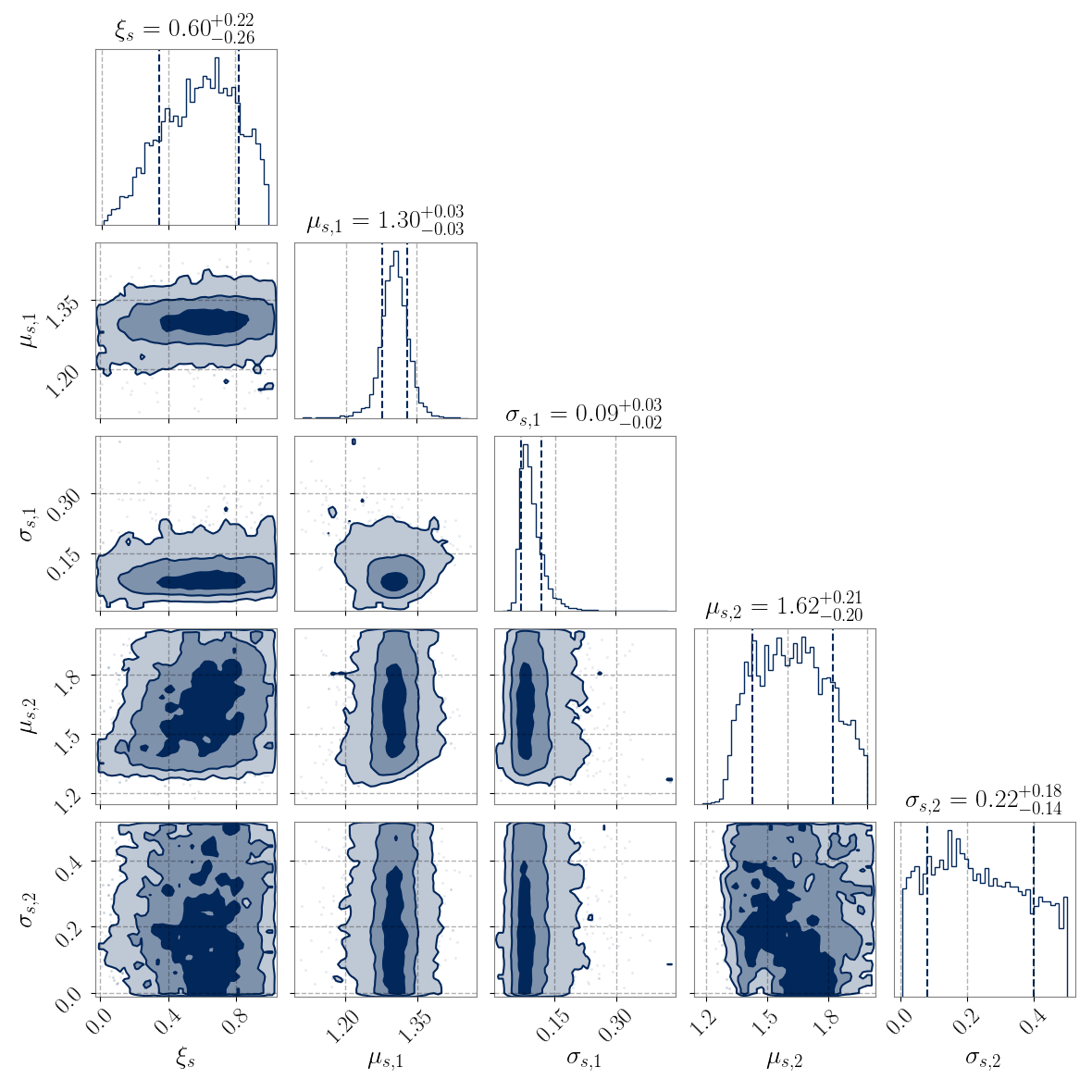}
    \caption{
    Posterior distributions for the slow mass population parameters.
    The credible intervals on the 2D posterior distributions are at $1\sigma$, $2\sigma$ and $3\sigma$, using increasingly light shading.
    The intervals on 1D posterior distributions are at $1\sigma$ level.
    }
    \label{fig:ms}
\end{figure*}

In Figs.~\ref{fig:mr}-\ref{fig:ms} we provide posterior corner plots for the recycled and slow population parameters.
The shaded regions indicate $1\sigma$, $2\sigma$, and $3\sigma$ credible intervals.
The high-mass peak in the mass distribution of slow neutron stars associated with the fast-merging channel visible in Fig.~\ref{fig:ppd_recycled_slow} is also shown in Fig~\ref{fig:ms}, which shows mild support for a broad peak at $\mu_{s2}\approx {1.62} M_{\odot}$.
We rule out the hypothesis that all DNS are from the high-mass peak at high credibility.
There is only mild support for the existence of a high-mass peak (Bayes factor ${\cal B}=2$).
While the presence of a fast-merging channel is slightly preferred $\xi_s<1$, the data are consistent with the hypothesis that no binaries merge swiftly $\xi_s=1$.

\begin{figure*}
    \includegraphics[width=\textwidth]{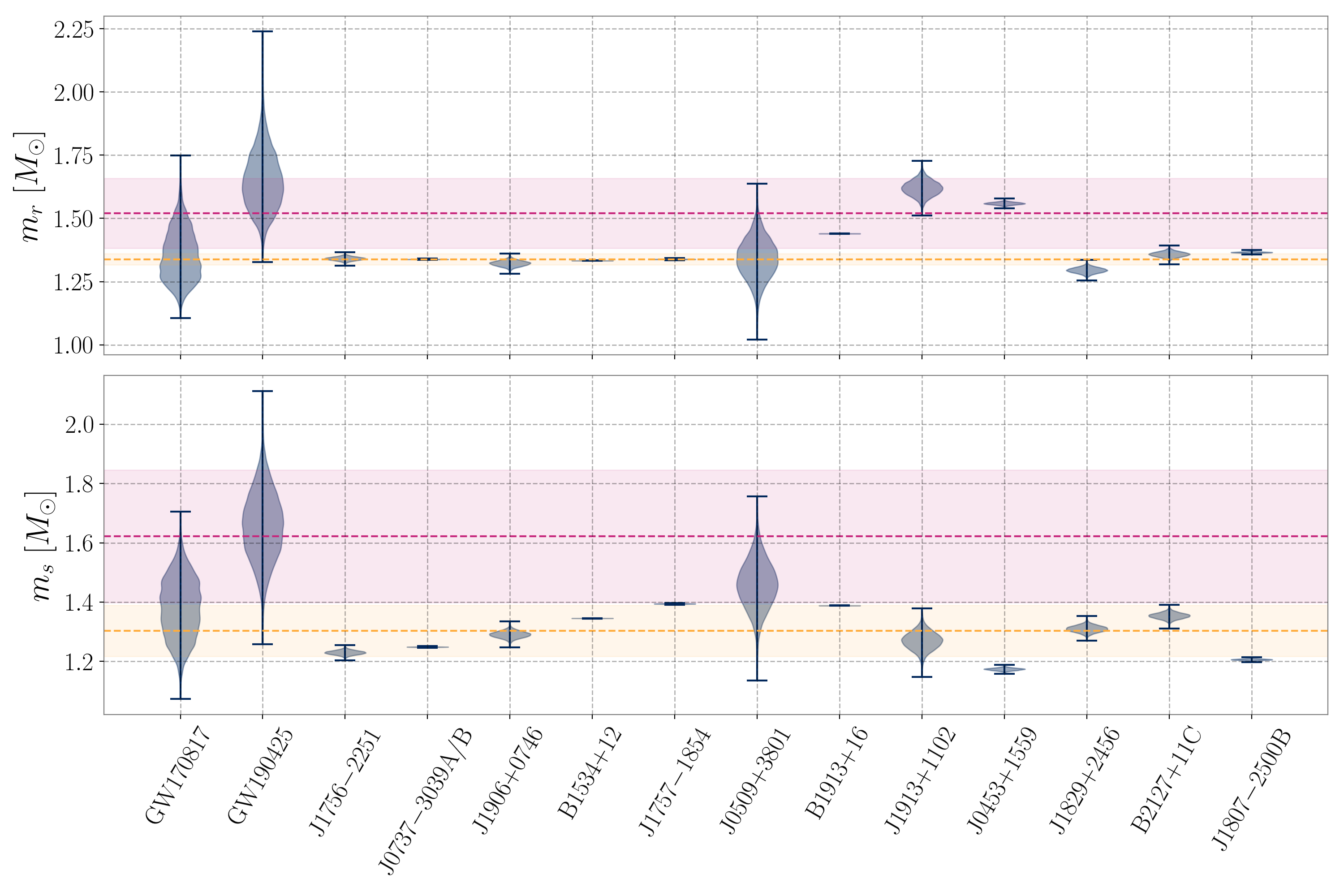}
    \caption{
    Posterior distributions of recycled and slow mass for the radio and gravitational-wave events.
    The shaded regions represent the 1$\sigma$ level of the low (yellow) and high (magenta) mass peaks in our mass distributions.
    Note that, for two globular-cluster pulsars (B2127+11C and J1807$-$2500B; both recycled), their undetected companions could be either young/slow or recycled neutron stars. The recycled and slow posteriors for the gravitational-wave events are from \cite{Zhu2020}. 
    }
    \label{fig:binary_posteriors}
\end{figure*}

\begin{figure*}
    \centering
    \includegraphics[width=0.55\textwidth]{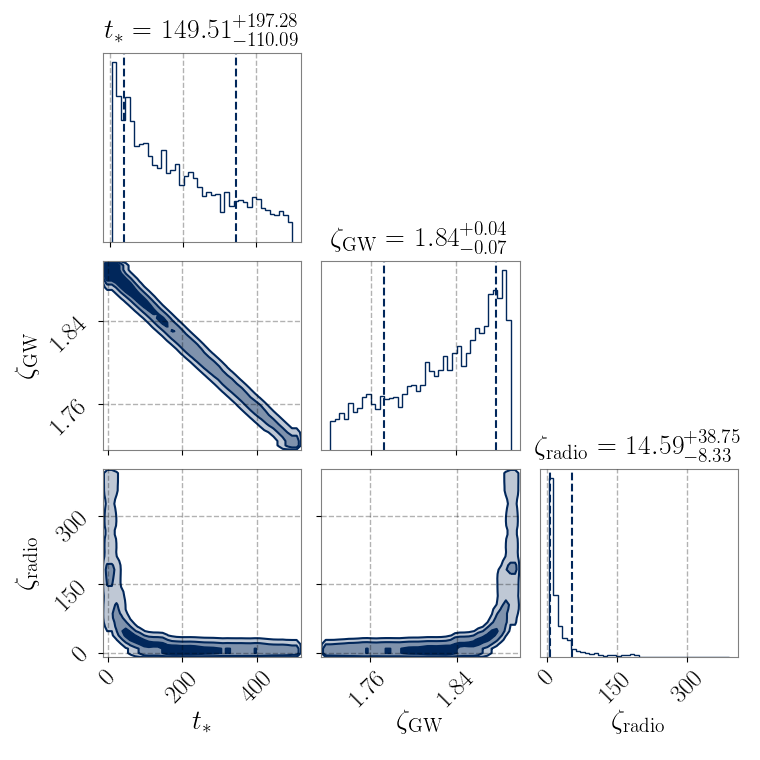}
    \caption{Posterior distributions for the characteristic delay time and transfer function weights.
    The credible intervals on the 2D posterior distributions are at $1\sigma$, $2\sigma$ and $3\sigma$, using increasingly light shading.
    The intervals on the 1D posterior distributions are at $1\sigma$.
    }
    \label{fig:transfer_posteriors}
\end{figure*}

In Fig.~\ref{fig:binary_posteriors}, we plot the mass posteriors for the DNS used in our study. 
The shaded regions indicate the 1$\sigma$ level of the low (yellow) and high (magenta) mass peaks from our population analyses.
The fast-merging channel is represented by the high-mass peak in the mass distribution of slow neutron stars ({$\mu_{s2}=1.62^{+0.31}_{-0.27}$} at 90\% credibility).
The heavy DNS merger, GW190425, is associated with this high-mass, fast-merging channel.
In Fig.~\ref{fig:transfer_posteriors}, we plot the posteriors for the characteristic delay time distribution and the corresponding distribution for the transfer functions: $\zeta_\text{GW}, \zeta_\text{radio}$. 
The shaded regions indicate $1\sigma$, $2\sigma$, and $3\sigma$ credible intervals.
We place only weak constraints on the delay-time parameter: {$t_{*}\approx 5-401$} (90\% credibility).
Each value of $t_{*}$ corresponds to single values of $\zeta_\text{GW}$ and $\zeta_\text{radio}$.
We find that {$\zeta_\text{GW} \approx 1.8$} regardless of the value of $t_{*}$ and $\zeta_\text{radio}$ ranges from {$\approx 4-74$} at 90\% credibility.

\section{Discussion}\label{sec:discussion}
Double neutron star systems evolve over time: they are born, they live for a period during which they may be seen in radio, and then they merge emitting gravitational waves.
It is necessary to model this evolution in order to achieve a clear understanding of DNS systems.
We present a model for the DNS birth mass distribution, along with a prescription for DNS evolution, which allows us to combine radio and gravitational-wave observations.
Using this framework, we are able to self-consistently measure the population properties of DNS at birth, at midlife (in radio), and at death (in gravitational waves).
It also enables us to eschew the conventional primary/secondary distinction of gravitational-wave astronomy in favour of an arguably more physically motivated description involving slow and recycled neutron stars as in \cite{Zhu2020}.\footnote{See also~\cite{Biscoveanu2020}, which models compact objects $A$ and $B$ depending on which one is more rapidly spinning.}

Hypothesising that the most massive DNS merge quickly (in $\sim 10-100~\mathrm{Myr}$) through unstable case-BB mass transfer, we find mild evidence for a broad peak at $\sim {1.6}~{M_{\odot}}$ in the birth mass distributions of slow neutron stars (see Fig.~\ref{fig:ms}).
We find mild support for the hypothesis that the unusually massive DNS system GW190425 can be understood as such a fast-merging system (${\cal B}=2$), which is consistent with the constraint on the merger time based on the spin measurement reported in \citet{Zhu2020}.
If we fix the delay-time of the fast merging channel to $t_*=10~\mathrm{Myr}$ we find slightly more support for this fast-merging scenario (${\cal B}=4$).
We find that {$\approx 8-79\%$} of binaries at birth are fast-merging.
If subsequent detections reveal a stronger discrepancy between the radio and gravitational-wave populations, our model provides a natural explanation for why such massive DNS are not commonly observed in radio since we infer that fast-merging DNS are suppressed in the radio population by a factor of {$\approx 4 - 74$ (90\% credibility)}.
Our model makes potentially falsifiable predictions about the maximum delay time for fast-merging DNS {$t_*\approx 5-401~\mathrm{Myr}$} (90\% credibility) and for the prevalence of fast-merging DNS in the Milky Way.

While our study shows that the presence of a fast-merging channel for massive DNS can account for a discrepancy between the Galactic population and the unusually large mass of GW190425, we do not rule out other explanations for the lack of heavy DNS visible in radio.
For example, \cite{Safarzadeh2020} argue that massive DNS are relatively common in the Milky Way, but we do not see them in radio because the magnetic field is buried through accretion.
\cite{Safarzadeh2020} argue that the fast-merging hypothesis explored here is in tension with at least some population synthesis studies, which predict a relatively small fraction of fast-merging systems: $\lesssim 10\%$.
This estimate is consistent with the low end of our 90\% credible interval: $8-79\%$.
However, theoretical predictions may vary with different population synthesis prescriptions.

It seems challenging to determine, which of these two scenarios is more likely to be true observationally since the buried-field hypothesis produces similar predictions to the fast-merging hypothesis.
The LISA observatory~\citep{LISA} may be able to provide evidence in favour of one hypothesis over the other; if high-mass DNS are plentiful in the Milky way, they should be visible in millihertz gravitational waves, even if they are radio-quiet \citep[][]{Lau:2019wzw,Andrews:2019plw,Korol:2020irk}.
Alternatively, if radio observatories are able to detect an ultra-relativistic DNS with a merger time of $\lesssim 10~\mathrm{Myr}$~\citep{Pol2020}, and the system includes an unusually massive neutron star, this would provide support for the fast-merging hypothesis.\footnote{If the buried-field approach turns out to be right, the framework presented in this paper can be straightforwardly adapted to describe that model instead.
For example, the fast-merging, high-mass peak would be reinterpreted as the buried-field, high-mass peak.
}
Finally, if radio observatories detect a high-mass DNS system with a long time to merge, such an observation could provide evidence against both the fast-merging hypothesis \textit{and} the buried-field hypothesis since one would expect such systems to be rare given either hypothesis.

While the analysis here is likely good enough to apply to the current small dataset of DNS, there are a number of improvements that will be useful to explore in the future as additional DNS are detected.
First, our model uses a simplistic delta-function delay-time distribution for the fast-merging channel.
With more DNS, the analysis will become more sensitive to the shape of the delay-time distribution, and it may become necessary to explore more realistic delay-time models.
Second, we ignore the possibility that the magnetic field might decay with time \citep{AshleyYuri17}.
A more sophisticated treatment could model this field decay.
Third, we assume that all DNS systems are equally detectable in the radio.
While this is probably a good approximation for our present purposes, it might be interesting to modify the analysis here to take into account the difficulty detecting ultra-relativistic DNS with radio~\citep{Pol2020}.
Finally, since all currently known DNS are relatively nearby, we carry out calculations using a zero-redshift approximation.
As LIGO--Virgo detect DNS mergers at ever greater distances, it will become possible to observe the changing proportions of slow and fast mergers over cosmic time.
Fortunately, we can straightforwardly accommodate this effect by making the radio and gravitational-wave transfer functions redshift-dependent.

Our results suggest that a reasonably large fraction ($\sim 40\%$) of DNS may merge quickly.
If so, the short merger time-scales of these systems have a number of implications.
Detections of short gamma-ray bursts are typically offset from their host galaxies and in regions where there is little star formation or in globular clusters \citep{Fong2013, Grindlay2016}.
Although these are fast-merging systems, these tight binaries ($P_\mathrm{orb} < 1~\mathrm{hr}$) can have high kick velocities, which can result in the binary escaping the host galaxy \citep{AndrewsZezas2019}.
If some DNS are found to merge quickly, future work should check for consistency with short gamma ray burst offsets.
Short merging timescales may also help explain $r$-process enrichment in globular clusters and ultra-faint dwarf galaxies \citep{Ji2016,Safarzadeh:2018fdy,Kobayashi2020}.
However, it is not clear that they will provide a significant source of the enrichment given the ejected mass from neutron star binaries decreases as the binary mass increases \citep{Radice2018}.
We are hopeful that this work will be useful when characterising the neutron-star black-hole mass gap.
By combining the DNS mass models here with models for the binary black hole mass spectrum, e.g.,~\cite{Talbot2018,o2_pop}, it should be possible to characterise the nature of this gap; see, e.g.,~\cite{Fishbach2020}.
Precise measurement of the mass gap may shed light on the minimum and maximum neutron star mass \citep{Chatziioannou2020}---or, at least, the minimum and maximum mass of neutron stars in DNS.
The maximum neutron-star mass has implications for the neutron star equation of state (EoS), which is determined by nuclear physics at the highest possible densities \citep{Breu2016,Essick2020,Landry2020}.

\section*{Acknowledgements}
We are thankful to Reed Essick for helpful discussions and feedback on early drafts of this paper.
We also thank Nikhil Sarin and Ryosuke Hirai for their insightful discussions and comments on this work.
This work is supported through Australian Research Council (ARC) Future Fellowship FT150100281 and Centre of Excellence CE170100004.
The authors are grateful for computational resources provided by the LIGO Laboratory and supported by National Science Foundation Grants PHY-0757058 and PHY-0823459.

\bibliographystyle{aasjournal}
\bibliography{main}

\appendix\label{appendix}
In this appendix we investigate the distribution of supernova kicks required for the fast-merging hypothesis of GW190425 to have explanatory power.
We carry out a rudimentary population synthesis analysis in order to estimate the typical supernova kick velocities required in order to disrupt a non-negligible fraction of slow-merging binaries (thereby changing the radio-visible mass distribution) without disrupting the fast-merging binaries created by unstable case-BB mass transfer.
While some population synthesis analyses assume maximum supernova kicks of $\sim\unit[1000]{km\, s^{-1}}$, we find that kicks of $\gtrsim\unit[2000]{km\, s^{-1}}$ are required for the fast-merging hypothesis to have explanatory power.
While such large kicks are potentially surprising, the existence of large kicks in at least some cases is supported by observations of high-velocity pulsars \citep{Hobbs05,Chatterjee:2005mj,Tomsick:2012ApJ}; we return to the topic of these high-velocity pulsars below.

The first step in our analysis is to assume a relationship between the progenitor carbon-oxygen (CO) core mass $m_\text{core}$ and the remnant mass $m$.
Different prescriptions assume different relationships.
The general trend is for the $m_\text{core}$ to increase with $m$~\citep{Spera,Fryer}, though, this increase may be interrupted by discontinuities~\citep{Mandel}.
Here, we assume a linear relationship between core mass and the mass of the slow neutron star:
\begin{align}\label{eq:}
    m_\mathrm{core} = 2.65 m_{s} - 0.95 .
\end{align}
As a rule of thumb, larger supernova kick velocities $v_\text{kick}$ are thought to be associated with larger core masses, although, it is possible there is significant stochasticity so that different kicks are possible for a single core mass.
We assume a linear relation
\begin{align}\label{eq:kick}
    v_\mathrm{kick} = km_\mathrm{core} .
\end{align}
We treat $k$ as a free parameter and vary it between $\unit[100-800]{km\, s^{-1} M_\odot^{-1}}$ in order to find a value that produces sufficiently large kicks to disrupt slow-merging massive binaries.
We simulate DNS systems, drawing masses from our mass distributions (Fig.~\ref{fig:ppd_recycled_slow}) and orbital periods from a log-uniform distribution. 
For the slow-merging channel, the orbital period $P_b$ range is $\unit[1-12]{hrs}$ \citep{Tauris17dns}; for the fast merging channel, the range is $\unit[0.5-1.5]{hr}$.
For each of DNS system we calculate $v_\text{kick}$ and compare it to the maximum kick that a binary can withstand without being disrupted $v_\text{max}$~\citep{Brandt}.
If $v_\text{kick}>v_\text{max}$, then the DNS is flagged as disrupted; otherwise it remains intact.

We carry out the simulation for different values of $k$.
For each value of $k$, we calculate the ``radio fraction'' $f_\text{radio}$, defined as the fraction of surviving slow-merging DNS with $P_b>\unit[1.5]{hr}$ and total mass greater than the total mass of GW190425:
\begin{align}\label{eq:radio_fraction}
    f_\text{radio} = \frac{N\Big(
    v_\text{kick} < v_\text{max}
    \,\&\,
    P_b > \unit[1.5]{hr}
    \,\&\,
    m_\text{tot}>m_\text{tot}^\text{GW190425}
    \Big)}
    {N\Big(
        v_\text{kick} < v_\text{max}
        \,\&\, 
        P_b > \unit[1.5]{hr}
    \Big)} .
\end{align}
The requirement that $P_b > \unit[1.5]{hr}$ is to take into account the fact that ultra-relativistic binaries are difficult to detect in radio~\citep{Pol2020}.
We expect $f_\text{radio}$ to become smaller for sufficiently large values of $k$ because slow-merging DNS are disrupted.
Next we calculate the ``GW fraction'' $f_\text{GW}$, defined as the fraction of surviving \textit{fast-merging} DNS with total mass greater than the total mass of GW190425:
\begin{align}\label{eq:gw_fraction}
    f_\text{GW} = \frac{N\Big(
    v_\text{kick} < v_\text{max}
    \,\&\,
    m_\text{tot}>m_\text{tot}^\text{GW190425}
    \Big)}
    {N\Big(
        v_\text{kick} < v_\text{max}
    \Big)} .
\end{align}

\begin{figure*}
    \centering
    \includegraphics[width=0.6\textwidth]{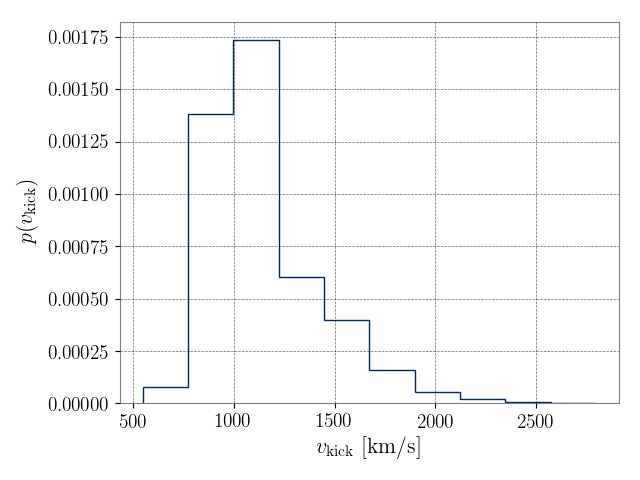}
    \caption{
    Histogram of kick velocity for slow-merging DNS assuming $k=\unit[400]{km\, s^{-1}M_\odot^{-1}}$ (see Eq.~\ref{eq:kick}).
    Given the assumptions of our model, this value of $k$ produces sufficiently large kicks for the fast-merging hypothesis to efficiently deplete massive, GW190425-like binaries from the radio population.
    Large kick values $\gtrsim\unit[1000]{km\, s^{-1}}$ pose a challenge to conventional thinking about supernovae.
    }
    \label{fig:kick-distribution}
\end{figure*}

The results are summarised in Table~\ref{tab:popsynth_slow_fast}.
For values of $k\lesssim\unit[300]{km\, s^{-1} M_\odot^{-1}}$, the fast-merging hypothesis has little explanatory power since the radio fraction is fairly stable at $f_\text{radio}\approx0.084$.
Approximately 8\% of the radio population are massive GW190425-like binaries, and these high-mass systems are not effectively disrupted by supernova kicks.
However, when $k\gtrsim\unit[400]{km\, s^{-1} M_\odot^{-1}}$, the radio fraction begins to fall appreciably, indicating that supernova are distorting the distribution of neutron star mass in radio-visible DNS.
Thus---given our simple model---we find that $k\gtrsim\unit[400]{km\, s^{-1} M_\odot^{-1}}$ is required in order for the fast-merging hypothesis to have explanatory power in accounting for the unusual mass of GW190425.
In Fig.~\ref{fig:kick-distribution}, we show the kick velocity distribution for $k=\unit[400]{km\, s^{-1}M_\odot^{-1}}$.\footnote{While a line with a slope of $\unit[400]{km\, s^{-1}M_\odot^{-1}}$ can be drawn through the $v_\text{kick}$-$m_\text{core}$ relation in \cite{Mandel}, we require larger values of $v_\text{kick}$ than allowed in that work in order for the fast-merging hypothesis to have explanatory power.}
We find that a significant fraction of neutron star kicks are greater than $\unit[1000]{km\, s^{-1}}$, which is sometimes taken as a plausible maximum value.
Some kicks in this model even exceed $\unit[2000]{km\, s^{-1}}$.
The exact shape of the supernova kick distribution required to support the fast-merging scenario is highly uncertain---a different parameterization could produce a substantially different result.
However, it seems clear that large kicks ($\gg\unit[1000]{km\, s^{-1}}$) are required for the fast-merging theory to have explanatory power.

Observations of pulsar proper motions indicate that neutron star kicks are typically a few hundred km\,s$^{-1}$ \citep[e.g.][]{Hobbs05}, although some pulsars are observed to have velocities greater than $\unit[1000]{km\, s^{-1}}$  \citep{Hobbs05,Chatterjee:2005mj,Tomsick:2012ApJ}.
\citet{Hobbs05} measure proper motions for PSRs B2011+38 and B2224+65, which imply both pulsars have velocities $> \unit[1500]{km\, s^{-1}}$. 
For PSR~B2224+65, this agrees with an independent velocity measurement based on the existence of a bow shock \citep[the so-called "guitar nebula";][]{Cordes:1993Nature}. 
PSR~B1508+55 has a well measured velocity of $> \unit[1000]{km\, s^{-1}}$ based on VLBA measurements of its proper motion and parallax \citep{Chatterjee:2005mj}.

A number of pulsars and central compact objects also have high inferred velocities ($> \unit[1000]{km\, s^{-1}}$) based on associations with supernova remnants (SNRs). 
PSR J1437$–$5959 has an inferred velocity $> \unit[1000]{km\, s^{-1}}$ based on an association with the SNR G315.78-0.23 \citep{Ng:2012ApJ}.
SGR 0525-66 has a velocity of $> \unit[1200]{km\, s^{-1}}$ based on association with SNR N49 \citep{Rothschild:1994Nature}, whilst the central compact object XMMU J172054.5$-$372652 in SNR G350.1-0.3 has an inferred velocity of 1400--2600\,km s$^{-1}$ \citep{Lovchinsky:2011ApJ}. 
Similarly, RX J0822-4300 in the SNR Puppis-A has a velocity of $\sim 750$--1500\,km s$^{-1}$ \citep{HuiBecker2006A&A}. 
IGR J11014-6103, associated with the SNR MSH 11–61A may be the fastest known pulsar with an inferred velocity of 2000--3000\,km s$^{-1}$ \citep{Tomsick:2012ApJ}.
Unfortunately, none of the neutron stars mentioned above have mass measurements (since typically mass measurements are only possible for neutron stars in binaries), so it is not possible to check if there is a correlation between large masses and high velocities.

For the fast-merging channel, $f_\text{GW}$ is fairly stable until $k\gtrsim\unit[700]{km\, s^{-1}}M_\odot$ at which point supernova kicks become so large that they begin to disrupt even tight, fast-merging binaries.
Interestingly, the entire range of $k$ considered in Table~\ref{tab:popsynth_slow_fast} is consistent with the observed mass distributions of DNS in radio and gravitational waves.
As noted above, small values of $k\lesssim\unit[300]{km\, s^{-1}}M_\odot$ do not appreciably affect the shape of the radio distribution of DNS.
However, since we find only mild statistical support for the fast-merging hypothesis, the radio data can be adequately explained by supposing that the radio DNS population contains massive GW190425-like systems like the gravitational-wave population---we just have not seen one yet.
This is not an unlikely possibility: given our model, the probability of not observing a radio DNS at least as massive as GW190425 from a set of 12 observations is 35\%.

It is also possible that large kicks are not required to explain the absence of massive DNS in radio observations. 
Perhaps the simplest explanation is that the massive helium stars invoked in the fast-merging hypothesis do not reach the giant radii of less massive He stars \citep[e.g.][]{Laplace:2020A&A}. 
If the maximum radius of these stars is only a few solar radii, then only binaries with $P_\mathrm{orb} \lesssim \unit[12]{hr}$ will interact. 
Therefore the first born neutron stars in wide massive DNS may be unrecycled and thus more difficult to observe in radio (due to their short radio lifetimes).
This hypothesis requires further investigating with detailed studies, and may be tested observationally with LISA~\citep{lisa_ska,Lau:2019wzw}.

\begin{table*}
\centering
\begin{tabular}{ccc} 
    \hline
    $k$ ($\unit[]{km\, s^{-1} M_\odot^{-1}}$)& $f_\text{radio}$ & $f_\text{GW}$ \\
    \hline\hline
    100 & 0.084 & 0.083\\
    200 & 0.084 & 0.083 \\
    300 & 0.082 & 0.083 \\
    400 & 0.059 & 0.083 \\
    500 & 0.042 & 0.083 \\
    600 & 0.02 & 0.081 \\
    700 & 0.004 & 0.06 \\
    800 & 0 & 0.034 \\
    \hline
\end{tabular}
\caption{
The relationship of ``radio fraction'' $f_\text{radio}$ (Eq.~\ref{eq:radio_fraction}) and the ``gravitational-wave fraction'' $f_\text{GW}$ (Eq.~\ref{eq:gw_fraction}) to $k$, the coefficient relating CO core mass and supernova kick (Eq.~\ref{eq:kick}).
The radio fraction is stable for small values of $k$.
However, when $k\gtrsim\unit[400]{km\, s^{-1}M_\odot^{-1}}$, $f_\text{radio}$ begins to drop appreciably, indicating that supernova kicks are efficiently disrupting slow-merging, massive DNS like GW190425 from the radio population.
The gravitational-wave fraction is stable until $k\gtrsim\unit[700]{km\, s^{-1} M_\odot^{-1}}$ at which point supernova kicks become so large that even tight, fast-merging binaries are disrupted, thereby depleting the GW190425-like DNS from the population observed with gravitational waves.
}
\label{tab:popsynth_slow_fast}
\end{table*}

\label{lastpage}
\end{document}